\documentclass[twocolumn]{aastex631}

\usepackage{units}

\usepackage{amsmath}

\newcommand{\code}[1]{\texttt{#1}}
\newcommand{\mesa}{\code{MESA}}
\newcommand{\MESA}{\mesa}

\newcommand{\Msun}{\ensuremath{\mathrm{M}_\odot}}

\newcommand{\gcc}{\ensuremath{\mathrm{g\,cm^{-3}}}} % density units

\newcommand{\nuclei}[2]{\ensuremath{\mathrm{^{#1}#2}}}

\newcommand{\carbon}[1][12]{\nuclei{#1}{C}}

\newcommand{\oxygen}[1][16]{\nuclei{#1}{O}}

\newcommand{\neon}[1][20]{\nuclei{#1}{Ne}}
\newcommand{\sodium}[1][23]{\nuclei{#1}{Na}}
\newcommand{\magnesium}[1][24]{\nuclei{#1}{Mg}}
\newcommand{\aluminum}[1][27]{\nuclei{#1}{Al}}

\begin{document}

\author[0000-0002-4870-8855]{Josiah Schwab}
\affiliation{Department of Astronomy and Astrophysics, University of California, Santa Cruz, CA 95064, USA}
\correspondingauthor{Josiah Schwab}
\email{jwschwab@ucsc.edu}

\received{05 May 2021}
\revised{17 May 2021}
\accepted{18 May 2021}

\title{Cooling Models for the Most Massive White Dwarfs}

\begin{abstract}
  We present a set of ultramassive white dwarf models, focused on
  masses above $1.3\,\Msun$.  Given the uncertainties about the
  formation and compositions of such objects, we construct
  parameterized model sequences, guided by evolutionary calculations
  including both single star and double white dwarf merger formation
  channels.  We demonstrate that the cooling of objects with central
  densities in excess of $10^9\,\gcc$ is dominated by neutrino cooling
  via the Urca process in the first $\approx 100$ Myr after formation.
  Our models indicate that the recently discovered ultramassive white
  dwarf ZTF J190132.9+145808.7 is likely to have experienced this Urca-dominated
  cooling regime.  We also show that the high densities imply that
  diffusion is unlikely to significantly alter the core
  compositions of these objects before they crystallize.
\end{abstract}


\section{Overview}

The most massive white dwarfs (WDs) represent extremes of stellar
evolution.  If created directly though the formation of a massive
degenerate core (whether in a single star or in a binary), they
represent the final fate of stars just below the mass where
supernova explosions occur \citep[e.g.,][]{Miyaji1980,
  Tauris2015b, Doherty2017}.  Alternatively, such a massive WD may be
created through the coalescence of two lower mass WDs and so its
properties reflect the merger process and its aftermath
\citep[e.g.,][]{Schwab2021}.

As our sample of nearby WDs becomes increasing complete, thanks in
large part to Gaia \citep{GaiaDR2, GentileFusillo2019a}, a significant
number of white dwarfs with masses $\gtrsim 1.3\,\Msun$ are being
revealed and better described.
These include long-studied objects like GD 50
\citep{Bergeron1991, Gagne2018} and RE
J0317-853 \citep{Barstow1995, Ferrario1997, Kulebi2010} as well more
recently characterized objects like WD J183202.83+085636.24
\citep{Pshirkov2020} and ZTF J190132.9+145808.7 \citep{Caiazzo2021}.
In their analysis of the 100 pc sample from the Montreal White Dwarf Database,
\citet{Kilic2021} identify 25 WDs with masses $> 1.3\,\Msun$ (assuming H atmospheres and C/O cores).

The core composition of ultramassive WDs is usually assumed to be
O/Ne, as is found in typical single star stellar evolution calculations.
However, this has begun to be questioned as the observed WD population
on the Q-branch appears to align with the location of C/O crystallization well out
into the ultramassive WD regime \citep{Cheng2019, Tremblay2019a,
  Bauer2020, Camisassa2021}.
There are also theoretical suggestions that carbon burning can fail to
reach the center \citep[e.g.,][]{Denissenkov2013b} or that effects of
rotation and mass loss can lead models to avoid carbon ignition
entirely \citep{Althaus2021}.

Given their uncertain formation pathways and limitations on the input
physics (e.g., equation of state, atmosphere tables), detailed WD
cooling studies often do not consider masses $\gtrsim 1.3\,\Msun$
\citep[e.g.,][]{Camisassa2019, Bauer2020, Bedard2020}.  Evolving the
highest mass models also requires including additional physical effects.  In
WDs with masses $\gtrsim 1.33$ \Msun\, and correspondingly densities
$\gtrsim 10^9\,\rm g\,cm^{-3}$, neutrino cooling through the Urca
process \citep{Gamow1941} can outpace plasmon decay to become the
dominant neutrino cooling process \citep{Tsuruta1970}.

In this paper, we study the cooling properties of the most massive WDs
($\gtrsim 1.3\,\Msun$).  Section~\ref{sec:methods} describes how we
construct our WD models and Section~\ref{sec:results} shows their
cooling tracks.  In Section~\ref{sec:conclusions} we summarize and
conclude with a comparison to ZTF J190132.9+145808.7.

\section{Methods}
\label{sec:methods}

We use the stellar evolution code Modules for Experiments in Stellar
Astrophysics \citep[\MESA;][]{Paxton2011, Paxton2013, Paxton2015,
  Paxton2018, Paxton2019} to generate and evolve our WD models.  Our
input and output files are publicly accessible at \url{https://doi.org/10.5281/zenodo.5047378}.

MESA solves the stellar structure equations under the assumption of Newtonian gravity.
For these massive WDs, general relativistic (GR) corrections begin to enter (i.e., $GM/(c^2r) \sim 10^{-3}(M/1.35\,\Msun)(r/1800\,{\rm km})^{-1}$).  This will lead to a small differences between the models presented here and ones that solve the Tolman-Oppenheimer-Volkoff equation.  The cold WD models from F.X. Timmes%
\footnote{\url{http://cococubed.asu.edu/code_pages/coldwd.shtml}} provide a rough guide to the magnitude of these differences.
At a fixed central density of $10^9\,\gcc$ ($10^{10}\,\gcc$) models with GR are $0.006\,\Msun$ ($0.013\,\Msun$)
less massive than those without GR.  In both cases, the radii are the same to within 3 km.
Therefore, masses effectively inferred from the radii of our models may be systematically high by $\approx 0.01\,\Msun$.

\subsection{Parameterized white dwarf models}
\label{sec:models}

Generating a large set of ultramassive WD models of varying core
compositions through self-consistent evolutionary calculations is
beset with difficulties.  This requires following models through the
challenging thermally-pulsing asymptotic giant branch and affords
only indirect control of the WD mass and core composition.
Given the uncertainties in the formation pathways, we want to
create models with both C/O and O/Ne core compositions.

Therefore, we generate a set of parameterized WD models with masses 1.29-1.36 \Msun\ (with a spacing of 0.01 \Msun)
using the \MESA\ \texttt{wd\_builder} capability.%
\footnote{Available in \texttt{mesa-contrib}: \url{https://github.com/MESAHub/mesa-contrib/}}
This feature provides an alternate initial model builder that creates a WD of a given mass and chemical composition.
Our models assume initially homogeneous chemical compositions in the core.
The initial thermal structure is approximated as a degenerate isothermal core with a temperature $2.5 \times 10^8$ K and a radiative envelope. The initial model is not in thermal equilibrium and so there is an initial transient phase lasting for roughly the conduction timescale of the core ($\sim 10^5$ yr) that should be disregarded.

We pick our detailed initial core compositions by running representative single star calculations at solar composition \citep[$Z = 0.017$;][]{Grevesse1998} in \MESA\ r15140.
Our C/O core composition is based on the averaged core composition of
an $\approx 1.0\,\Msun$ C/O WD evolved from a $6.4\,\Msun$
single star using the test suite case
\texttt{make\_co\_wd} and the nuclear network
\texttt{mesa\_49.net}.
Our O/Ne core composition is based on the averaged core composition of the $\approx 1.2\,\Msun$ degenerate core of an $8.0\,\Msun$ single star halted at its first thermal pulse.  This calculation used the nuclear network
\texttt{sagb\_NeNa\_MgAl.net}.

% c12 3.7e-01
% o16 6.0e-01
% ne20 2.2e-03
% ne22 1.6e-02
% na23 2.5e-04
% mg24 5.2e-04
% mg25 3.0e-04
% mg26 4.2e-04
% si28 6.6e-04
% s32 3.1e-04
% s34 1.6e-03

% 0.8406287483594663
% neut 2.2e-03
% c12 6.7e-03
% o16 5.7e-01
% ne20 3.2e-01
% ne22 1.4e-02
% na23 6.0e-02
% mg24 2.6e-02
% mg25 1.2e-03
% mg26 1.5e-04
% al27 2.8e-03

\begin{table}
  \centering
  \caption{The set of compositions used in our parameterized WD
    models.  Each column lists the mass fractions of the isotopes that
    were included.  Small adjustments are made to the
    \oxygen[16] abundance to ensure the mass fractions sum to 1.}
  \label{tab:compositions}
  \begin{tabular}{r|DD|D}
    \hline
    Isotope & \multicolumn2c{C/O models} & \multicolumn2c{O/Ne models} & \multicolumn2c{Merger model} \\
    \hline
    \decimals
    \carbon[12]    & 0.37     & 0.0067 &  0.0018 \\
    \oxygen[16]    & 0.61     & 0.57   &  0.44 \\
    \neon[20]      & 0.0022   & 0.32   &  0.45 \\
    \neon[22]      & 0.016    & 0.014  &  0.0022 \\
    \sodium[23]    & 0.00025  & 0.060  &  0.030 \\
    \magnesium[24] & 0.00052  & 0.026  &  0.050 \\
    \magnesium[25] & 0.00030  & 0.0012 &  0.015 \\
    \aluminum[27]  & 0        & 0.0028 &  0.0068 \\
    \hline
  \end{tabular}
\end{table}

The nuclear networks used in the single star models were chosen to
cover the main hydrogen, helium, and carbon burning phases.  However,
we need not retain all of these isotopes during the WD cooling
calculations. For simplicity, our WD models will only consider the
most abundant isotopes \carbon[12]/\oxygen[16]/\neon[20] and the key
neutron rich isotopes that sediment and/or participate in the Urca
process.
Table~\ref{tab:compositions} shows the isotopes and abundances that
define the composition of our C/O and O/Ne WD cooling models.
On top of a homogeneous core, all our models have an added He layer of $\sim 10^{-5}\,\Msun$
and no H layer.

\subsection{Evolutionary merger model}

As a point of comparison for the compositions used in cooling models,
we generate a double WD merger model following \citet{Schwab2021}.
Reflecting the simplified compositions and nuclear networks adopted in
hydrodynamical merger calculations, that work considered merger models
with uniform initial compositions of 40 per cent \carbon[12] and 60
per cent \oxygen[16] by mass and followed the evolution using an
$\alpha$-chain nuclear network.  Here, we improve upon that by
beginning with a more detailed composition and using a larger nuclear
network.

We first generate a model of an typical $\approx 0.6\,\Msun$ C/O WD
evolved from a $3.1\,\Msun$ single star with a solar
composition \citep[$Z = 0.017$;][]{Grevesse1998} using the test suite
case \texttt{make\_co\_wd} in \MESA\ r15140 and the nuclear network
\texttt{mesa\_49.net}.

We continue to follow the assumption of a uniform initial condition,
neglecting possible variation in composition between the primary and
secondary WD.  The fate of the WD surface H/He layers is uncertain
during the initial mass transfer and merger process.  Because we are
eliding the merger itself, we have no information about how much H/He
survives and where it is located.  Therefore, we make the simplifying
choice to average core composition (excluding the H and He rich outer
layers) of the $\approx 0.6\,\Msun$ C/O WD and use this as the uniform
initial composition for the merger model.

To follow the nuclear burning (primarily carbon burning) that occurs
in the merger, we adopt the nuclear network \texttt{mesa\_49.net},
which covers carbon burning and includes the neutron-rich isotopes
\sodium[23] and \magnesium[25].  As our example, we generate and
evolve a $q = 0.9, M_{\rm tot} = 1.35\,\Msun$ merger model.  For
masses $\gtrsim 1.35\,\Msun$ the occurrence of Ne-burning means that
our merger models no longer produce O/Ne WDs.
Therefore, this represents approximately the most massive WD that we
can make within this framework.
We include a small amount of mass loss, which is necessary to allow
\MESA\ to evolve the model onto the cooling track, and this means the
resulting WD model is $1.344\,\Msun$.

The rightmost column of Table~\ref{tab:compositions} shows its
composition.  Relative to the single star O/Ne composition, the O/Ne
ratio in the merger is more \neon[20] rich.  The total amount of
neutron-rich isotopes is similar, though the distribution is different,
with less \neon[22] and more \magnesium[25] in the merger model.
Given its overall similarity to the O/Ne composition from
Section~\ref{sec:models}, we do not generate and cool a separate set
of parameterized models to represent mergers.

\subsection{White Dwarf Cooling}

We cool our WD models using the \MESA\ development version (commit 5e701e79).
This allows us to take advantage of the Skye equation of state (EOS)
\citep{Jermyn2021}.  This free-energy-based EOS self-consistently
determines the location of the liquid/solid phase transition and the
accompanying latent heat release in multicomponent plasmas.
MESA and Skye do not currently include phase separation, so the
solid and liquid compositions are assumed to be the same.
All isotopes listed in Table~\ref{tab:compositions} have sufficiently
large mass fractions that they are included in the Skye EOS calculation.
Regions with partial ionization are covered by the FreeEOS \citep{Irwin2004} and
SCVH \citep{Saumon1995} EOSes.

We include the effects of diffusion via the approach described in
\citet{Paxton2018} and using the diffusion coefficients of
\citet{Stanton2016}.  All isotopes included in the nuclear network
diffuse and, following the approach in \citet{Bauer2020}, all isotopes
contribute to the sedimentation heating.

During the cooling phase, the nuclear network includes only the
electron captures and beta decays linking the \sodium[23]-\neon[23],
\magnesium[25]-\sodium[25], and \aluminum[27]-\magnesium[27] Urca
pairs.  Doing so ignores the possibility of pyconuclear
fusion of \carbon[12] \citep[e.g.,][]{Yakovlev2006}, which could occur
in the highest mass models that reach
$\log(\rho/\rm g\,cm^{-3}) \gtrsim 9.5$.
We use the on-the-fly weak reaction framework described in
\citet{Schwab2015} and \citet{Paxton2015} as this approach continues
to provide accurate rates at temperatures below $10^8$ K where interpolation in published tables can lead to undercooling artifacts \citep[see Appendix D
in][]{Schwab2017a}.
Thermal neutrino loss rates are from \citet{Itoh1996}, except for
plasmon neutrinos where we use the more accurate rates from \citet{Kantor2007}.

We use the conductive opacities of \citet{Cassisi2007}, supplemented
by the \citet{Blouin2020a} revisions for H and He.
Radiative opacities are from OPAL \citep{Iglesias1993, Iglesias1996}.
However, conditions are often off these tables, and in that
circumstance the radiative opacity values are extrapolated from the
table boundaries at constant temperature.%
\footnote{
    The OPAL tables used in MESA are tabulated as functions of $\log T$ and
    $\log R \equiv \log(\rho/{\rm g\,cm^{-3}}) - 3 \log(T/{10^6\,\rm K})$.
    The relevant (high-$R$) edge of the tables occurs at $\log R$ = 1.
    For conditions corresponding to $\log R > 1$,
    MESA uses the radiative opacity from the OPAL table at
    the input value of $\log T$ and $\log R = 1$.
    This radiative opacity is then combined with the conductive opacity in the usual way.
    See Section 4.3 and Figures 2 \& 3 in \citet{Paxton2011}.
}
When the conductive opacity dominates (i.e., is smaller than the
radiative opacity), this does not significantly affect the total
opacity.  But as helium becomes neutral and the radiative opacity
falls, the lack of appropriate opacity tables becomes more
problematic.  Therefore, we stop our models when they reach
$T_{\rm eff} = 1.5 \times 10^4$ K.

We use the \MESA\ ML2 \citep{Bohm1971} version of mixing length theory
(MLT), with a mixing length $\alpha = 1.8$.  We completely deactivate
MLT in solid material.  We also suppress chemical mixing due to
convection while continuing to allow convection to transport energy
(change the temperature gradient).
The Urca-process cooling leads to a superadiabatic temperature
gradient between the Urca shell%
\footnote{The neutrino emissivity
  associated with the Urca process peaks over a narrow range in
  density \citep[e.g.,][]{Paczynski1973a} and hence the region of
  significant cooling is typically a spherical shell.  This is centered at
  the threshold density of the Urca pair, where the electron chemical
  potential is equal to the difference between the chemical potentials of
  the mother and daughter ions, and its width depends on the temperature.
  In this region, the equilibrium
  composition has significant abundances of both species, thereby
  allowing for repeated electron capture and beta decay
  and the corresponding neutrino production.}
and the uncooled material at higher densities
(and hence produces a convective region).  If convective mixing is
allowed to transport material across the Urca shell, \MESA\
experiences both physical and numerical difficulties.  These have
their origins in an MLT treatment that allows convection to
instantaneously switch on/off and which does not account for the
interactions of convection with the Urca shell
\citep[e.g.,][]{BisnovatyiKogan2001}. Our inability to accurately
model this process introduces some uncertainties in our cooling rates,
but when mixing is allowed to drive the composition further from its
weak equilibrium, even more rapid neutrino cooling will result.
Therefore, our choice to suppress this mixing should not lead us to
overestimate the importance of the Urca cooling effect.

The suppressed mixing also includes the surface layers, so even as the
convection zone deepens, the surface remains pure He.  While
unrealistic, this has two advantages.  First, the outer layers of our
models are not self-consistently generated, so their detailed
structure (e.g., thickness of He envelope, composition of layers that would be
created during the elided thermal pulses) seems unlikely to be
reliable.  Second, not allowing the surface to become polluted by
mixed up material (primarily $\carbon[12]$) simplifies the application
of the outer boundary condition by not requiring radiative opacities for
metal-rich mixtures.  Because our most massive WD models have
$\log g > 9.4$ and so are are off the DB WD atmosphere tables included in MESA,
we use an Eddington grey atmosphere.
Again, we halt our models when they reach  $T_{\rm eff} = 1.5 \times 10^4$ K,
reflecting our lack of appropriate radiative
opacities for cool He and that appropriate atmosphere tables would
become increasingly important at lower effective temperature.

To demonstrate that our WD cooling models are reasonable, we compare
to some of the most massive models in the literature.  We do not
attempt to match WD properties or input physics assumptions, so these
comparisons are not expected to yield precise agreement.

\begin{figure}
  \centering
  \includegraphics[width=\columnwidth]{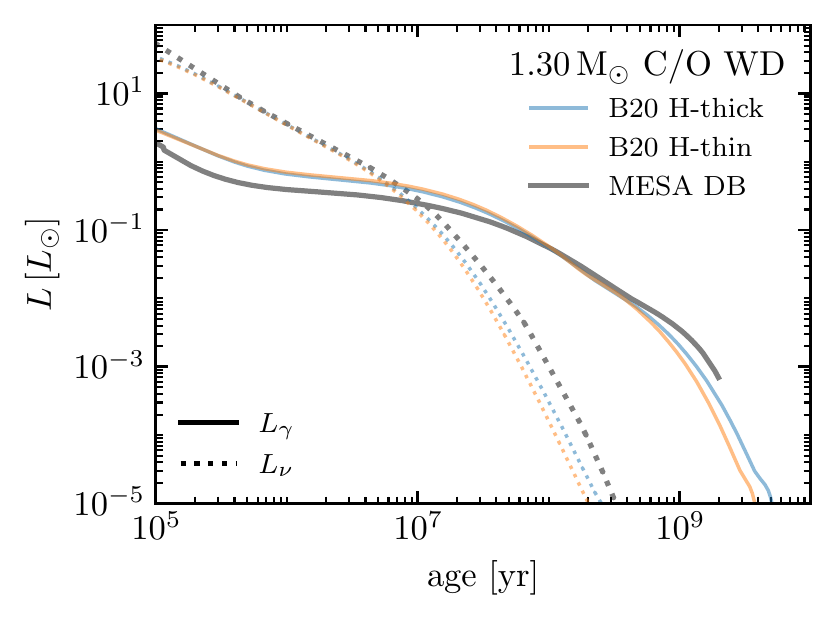}
  \caption{Luminosities of a cooling 1.30 \Msun\ C/O WD.  The \MESA\ model from this work is compared to the results of \citet{Bedard2020}.}
  \label{fig:B20}
\end{figure}

Figure~\ref{fig:B20} compares to the most massive C/O core model of
\citet{Bedard2020} which has a mass of $1.30\,\Msun$.  The He layer in
this model has a fractional mass $q_{\rm He}=10^{-2}$ and there is both a thin H layer
model ($q_{\rm H}=10^{-10}$) and a thick H-layer model
$(q_{\rm H}=10^{–4})$.  The interior composition of the
\citet{Bedard2020} is 50/50 C/O, resulting in a higher specific heat
then our roughly 40/60 C/O models.
Agreement is qualitatively good, always within a factor of $\approx 2$ in luminosity
at constant age.

\begin{figure}
  \centering
  \includegraphics[width=\columnwidth]{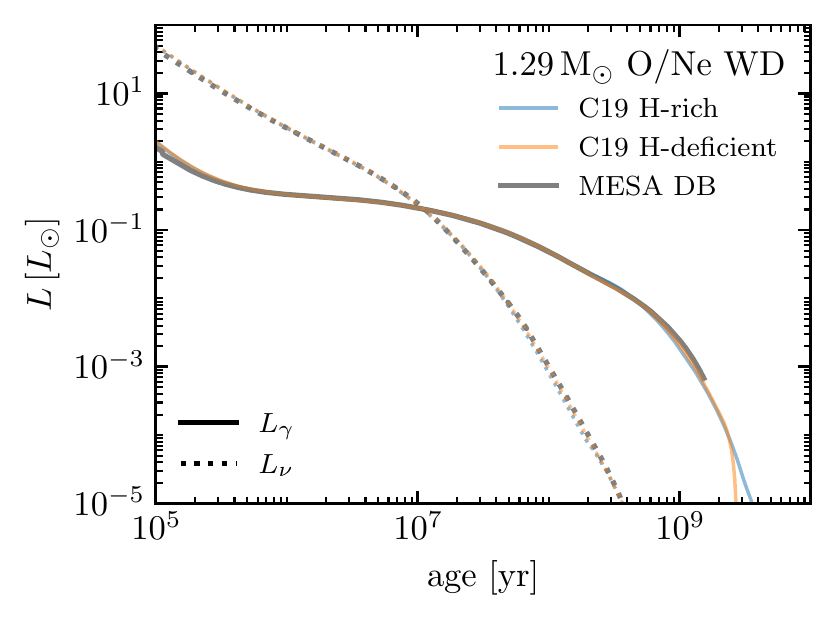}
  \caption{Luminosities of a cooling 1.29 \Msun\ O/Ne WD.  The \MESA\ model from this work is compared to the results of \citet{Camisassa2019}.}
  \label{fig:C19}
\end{figure}

Figure~\ref{fig:C19} compares to the most massive O/Ne core model of
\citet{Camisassa2019} which has a mass of 1.29 \Msun.  The mass of the
He layer in this model is $\approx 2\times10^{-5}\,\Msun$ and there
is both a H-deficient model and (no H layer) and a
H-rich model ($\sim 10^{-6}\,\Msun$ H layer).
The core chemical composition of our models are in rough agreement.
We do not include energy from phase separation upon crystallization.
Agreement is quantitatively good with agreement to within the line
width ($\sim 10$ \%) persisting out until ages of about 1 Gyr.

The agreement between the cooling tracks from \citet{Bedard2020} and
\citet{Camisassa2019} with varied envelope properties shown in
Figures~\ref{fig:B20}~and~\ref{fig:C19} also serve to illustrate that
surface composition (and use of atmosphere tables for the outer
boundary condition) has a relatively minor effect until later times
($\gtrsim 1$ Gyr), beyond the time of primary interest in our study.

\section{Results}
\label{sec:results}

\citet{Tsuruta1970} used a $1.373\,\Msun$ WD model to show that
Urca-process cooling is the dominant neutrino cooling mechanism
for $T \lesssim 2\times10^9\rm\,K$.  For the \sodium[23]-\neon[23] Urca
pair to operate, the central density of the WD must be above the
threshold density $\rho \gtrsim 1.6\times10^9\,\rm g\,cm^{-3}$.  At
these densities and temperatures the dominant non-nuclear neutrino
cooling processes is plasmon decay, followed by electron-ion bremsstrahlung \citep[e.g.,][]{Winget2004}.
The dominance of plasmon neutrinos motivated our choice to override the \MESA\ default rates from \citet{Itoh1996} with the more accurate rates from \citet{Kantor2007}.
The rate of Urca-process neutrino cooling scales linearly with the mass fraction of the Urca pair
and so the Urca process will be more important in O/Ne WDs than C/O WDs because of their greater \sodium[23]-\neon[23] and \magnesium[25]-\sodium[25] abundances.

\begin{figure}
  \centering
  \includegraphics[width=\columnwidth]{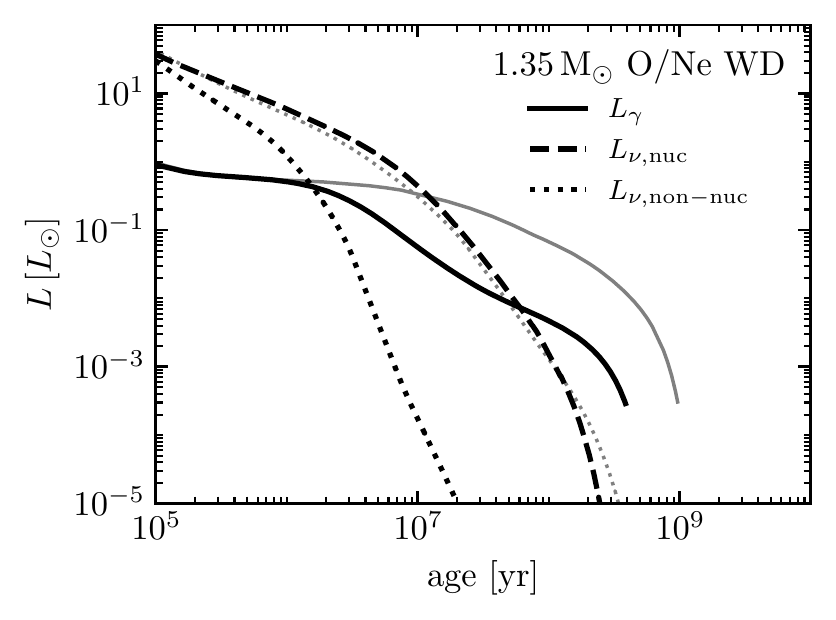}
  \caption{Comparison between models of a 1.35 \Msun\ O/Ne WD with (black) and without (grey) Urca-process neutrino cooling.  The solid lines show the surface (photon) luminosity, the dashed lines show the Urca-process neutrino luminosity, and the dotted lines show the non-nuclear (thermal) neutrino luminosity.}
  \label{fig:Urca_ONe}
\end{figure}

\begin{figure}
  \centering
  \includegraphics[width=\columnwidth]{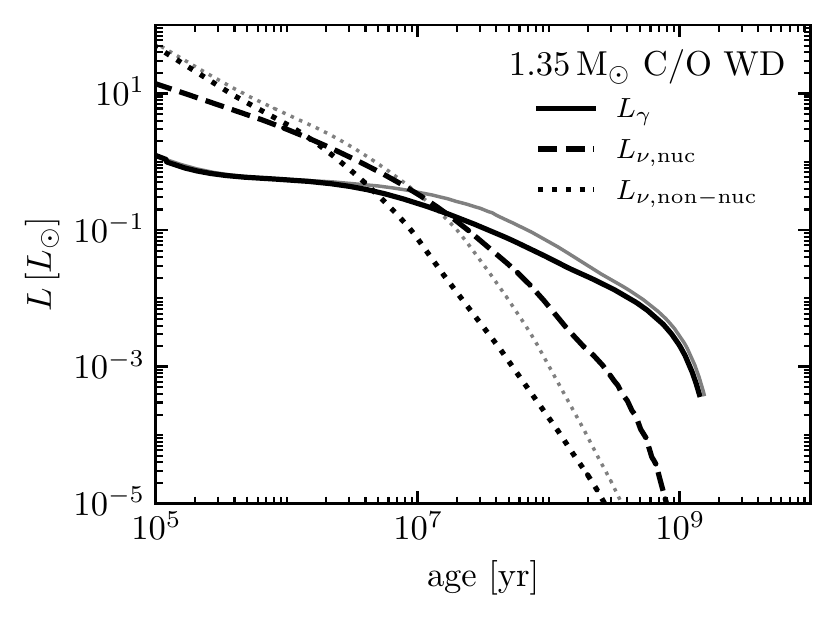}
  \caption{Comparison between models of a 1.35 \Msun\ C/O WD with and without Urca-process neutrino cooling.  Lines have the same meaning as in Figure~\ref{fig:Urca_ONe}.}
  \label{fig:Urca_CO}
\end{figure}

Figure~\ref{fig:Urca_ONe} compares the luminosity evolution of a $1.35\,\Msun$
O/Ne WD with and without the inclusion of Urca-process neutrino
cooling.  The model with Urca-process cooling experiences a
significant drop in luminosity around 10 Myr and beyond that point is
roughly a factor of ten less luminous at constant age, reaching our
$T_{\rm eff} = 1.5\times10^4$ K stopping condition in about half the total time
($\approx 600$ Myr less).
The dashed black line shows the Urca neutrino cooling rate, which is
always greater than the non-nuclear neutrino cooling rate shown by the
black dotted line.  Because this additional cooling leads to a lower
temperature, it reduces the loss rates through the other non-nuclear
mechanisms, and as such becomes the new dominant source rather than
acting as an additional source in an additive manner.
Thus comparing to the calculation without Urca-process cooling---its non-nuclear neutrino loss rate is shown by the dotted grey line---we see that the total neutrino cooling rate is less dramatically
different between the calculations, though still altered enough to yield a significant change in the cooling.

Figure~\ref{fig:Urca_CO} compares the luminosity evolution of a
$1.35\,\Msun$ C/O WD.  The lower Urca pair abundances lead to a less dramatic
effect, with the luminosity lower by a factor of $\approx 2$ at
constant age between 10 and 100 Myr.  The overall difference in the
cooling time to reach the $T_{\rm eff} = 1.5\times10^4$ K stopping condition
is $\approx 100$ Myr.

\begin{figure*}
  \centering
  \includegraphics[width=\columnwidth]{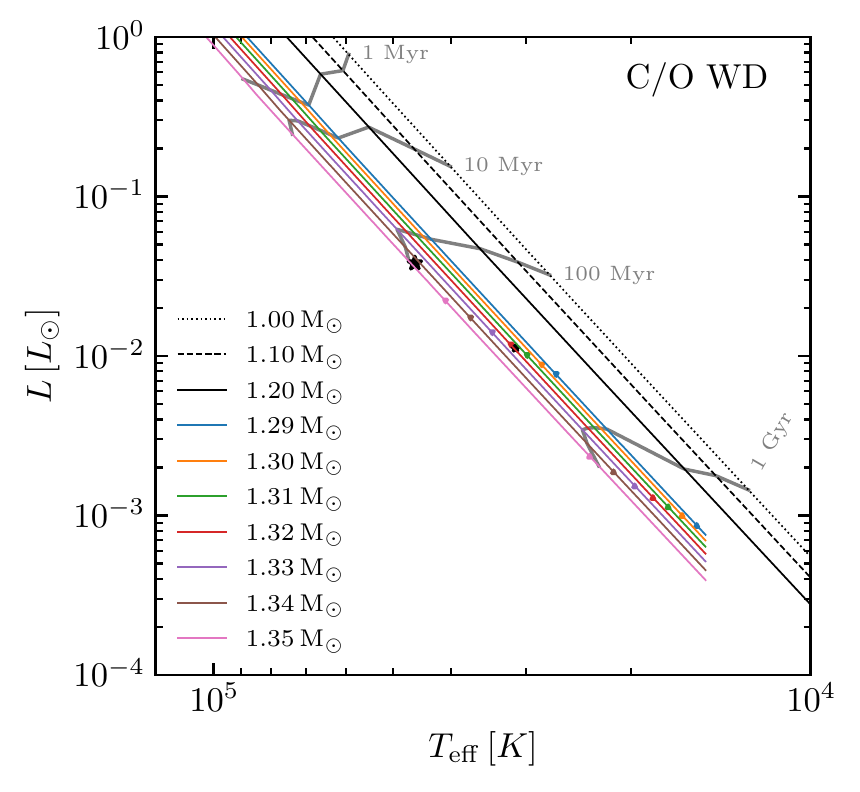}
  \includegraphics[width=\columnwidth]{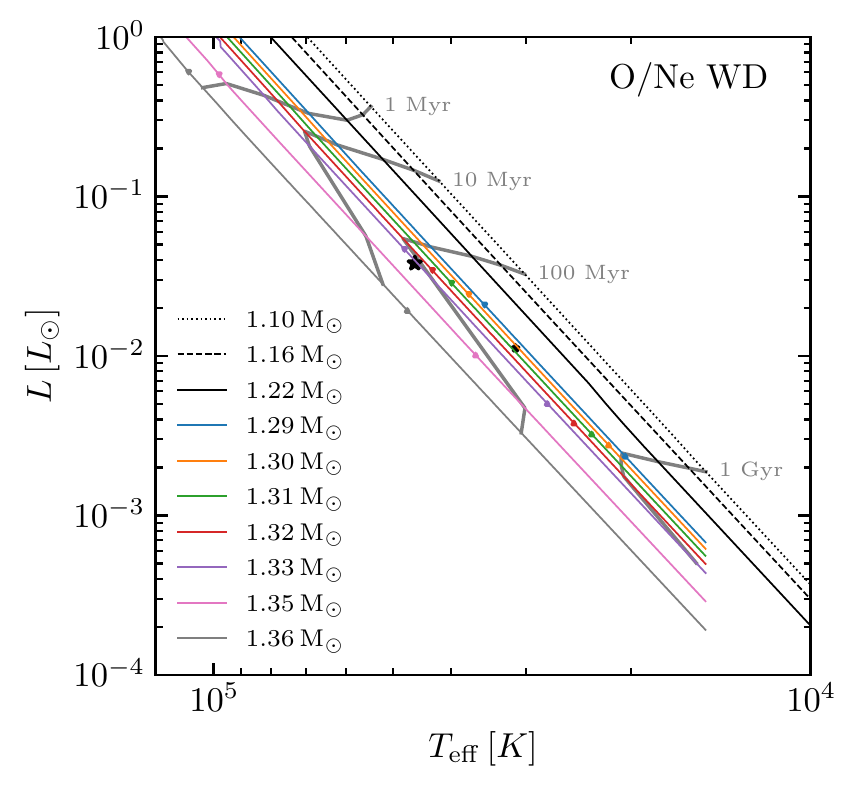}
  \caption{HR diagrams of C/O WDs (left panel) and O/Ne WDs (right panel). The thin lines solid lines are the WD cooling curves.  The solid grey lines are the labeled isochrones.  The black lines are models from the literature: C/O models are from \citet{Bedard2020}; O/Ne models are from \citet{Camisassa2019}.  The colored lines are models from this work.  The downward triangles mark when 10 \% of the mass is solid while the upward triangles mark when 90\% of the mass is solid.  The large star shows ZTF J190132.9+145808.7 \citep{Caiazzo2021} and the small star shows WD J183202.83+085636.24 \citep{Pshirkov2020}.}
  \label{fig:HR}
\end{figure*}

Figure~\ref{fig:HR} shows HR diagrams for our C/O WD (left panel) and
O/Ne WD (right panel) cooling sequences.  The
triangles on the cooling tracks mark the interval when the core is
crystallizing (starting when 10 \% of the mass is solid and ending
when 90 \% of the mass is solid).  Driven by the higher interior
densities, higher mass WDs crystallize at younger ages and higher
luminosities than lower mass WDs.
As an example, the $1.32\,\Msun$ C/O WD begins crystallizing at
cooling age $\approx 400$ Myr and finishes around $\approx 1.6$ Gyr.
Reflecting their higher charge, the $1.32\,\Msun$ O/Ne WD begins
crystallizing at cooling age $\approx 160$ Myr and finishes around
$\approx 800$ Gyr.

At the masses where Urca cooling becomes active
($\gtrsim 1.33\,\Msun$), the acceleration in the cooling is apparent
in the downward bend of the isochrones.
In the case of the C/O WDs, the luminosities at which crystallization
occurs smoothly continues its trend, but for the O/Ne WDs,
crystallization begins to occur at cooling ages below 1 Myr.
This happens because of the strong cooling at the location of the Urca
shell and the continued outward transport of heat from regions
interior to the shell.  This results in the rapid cooling and
crystallization of the material interior to the Urca shell.

The timing and physical extent of crystallization has important
implications for the cooling of WD models.
In addition to controlling when the latent heat of the phase
transition is released, once material becomes solid, the gravitational
potential energy stored in neutron-rich isotopes can no longer be
released through their preferential transport towards the center.
Realizing a scenario where a massive WD is meta-stable and will exceed
its effective Chandrasekhar mass due to sedimentation, such as is
speculated by \cite{Caiazzo2021} for ZTF J190132.9+145808.7, requires
rapid and near-complete sedimentation.

\begin{figure}
  \centering
  \includegraphics[width=\columnwidth]{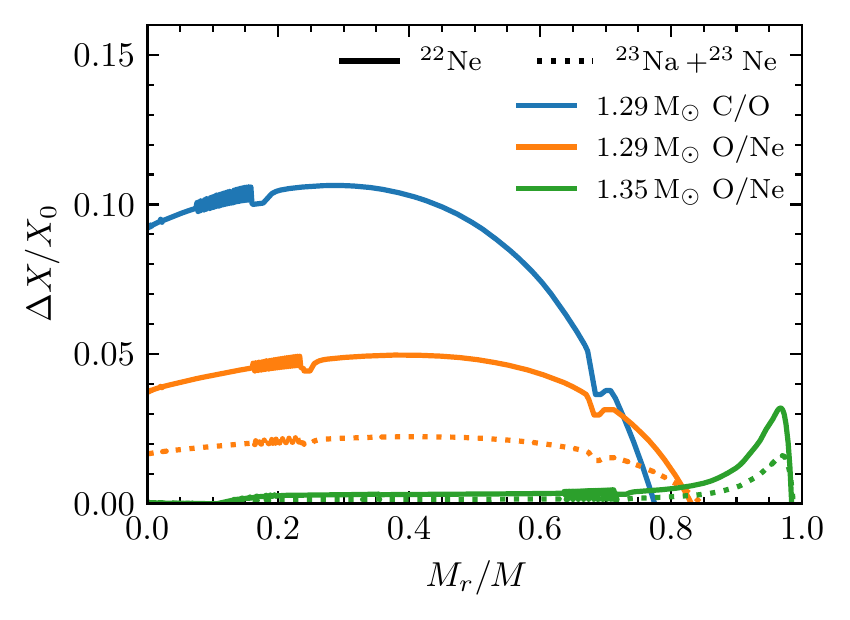}
  \caption{Fractional change in indicated neutron-rich isotopes in selected C/O
    WD and O/Ne WD models over the duration of our calculations.
    At their end, the WD cores are solid, inhibiting further change.}
  \label{fig:abundances}
\end{figure}

Figure~\ref{fig:abundances} shows the relative change in the neutron
rich abundances realized in the WD interiors due to gravitational
sedimentation.
This is the difference between the chemically 
    homogeneous initial condition and the chemical profile when the model reaches the
    $T_{\rm eff} = 1.5\times10^4{\rm\,K}$ termination condition.  At that time the models are nearly completely crystallized, with the 1.29 \Msun\, C/O WD,
  1.29 \Msun\, O/Ne WD, and 1.35 \Msun\, O/Ne WD models respectively having only $1.1 \times 10^{-1}\,\Msun$,
$1.8 \times 10^{-2}\,\Msun$, and $5.3\times10^{-4}\,\Msun$ of non-solid material.
At 1.29 \Msun\, the C/O WD shows a larger change than
the O/Ne WD, primarily reflecting its longer time to crystallization.
The 1.35 \Msun\ O/Ne shows almost no sedimentation in the interior,
reflecting its rapid crystallization due to Urca-process cooling.

\citet{Cheng2019} report a population of ultramassive WDs on the
Q-branch that experience multi-Gyr cooling delays (relative to both
current models and other observed WDs).  If explained by energy
release from sedimentation of neutron-rich isotopes, this requires
that effective diffusion coefficients for these species are
significantly increased over the single-particle results.
As discussed in \citet{Bauer2020}, the \citet{Stanton2016} diffusion
coefficients are in agreement at the factor of $\approx 2$ level with
the molecular dynamics results of \citet{Hughto2010}, and so this can
not be easily explained as a theoretical uncertainty in the
single-particle diffusion coefficients.
This led \citet{Bauer2020} to suggest that the formation of \neon[22]
clusters might accelerate sedimentation.  However, \citet{Caplan2020c}
suggest such clusters are not stable.
The enhancements invoked by \citet{Bauer2020} would still be far below
those necessary to achieve near-complete sedimentation.

Very recently, \citet{Blouin2021} calculated detailed phase curves for
C/O/Ne mixtures and suggested that a phase-separation-driven
``distillation'' process can occur.
Because this process occurs though buoyant rising of \neon[22]-deficient solid crystals,
  the net inward transport of neutron-rich material is not limited by the rate of
  single-particle diffusion.  The transport will eventually halt once the core composition
  reaches a point where there is no more phase separation and then the core solidifies.%
\footnote{This threshold will depend on the detailed phase diagram but
  for C/O/Ne, \citet{Blouin2021} find this is a \neon[22] number
  fraction of 20\%.} %
Future work should investigate the role of phase
separation in the most massive WDs.

\section{Conclusions}
\label{sec:conclusions}

We present models of WDs with masses $\gtrsim 1.3\,\Msun$ and with C/O
or O/Ne cores.  Our models are in good agreement with existing models
from \cite{Bedard2020} and \cite{Camisassa2019} at masses
$\approx 1.3\,\Msun$.  We include the results of Urca-process neutrino
cooling and demonstrate the significant acceleration of cooling that
occurs at masses above $1.33\,\Msun$, especially for O/Ne cores.

As an example, we apply our models to ZTF J190132.9+145808.7 \citep{Caiazzo2021}.
This WD is measured to have
$T_{\rm eff} = 46^{+19}_{-8}\,\rm kK$ and radius
$2140^{+160}_{-230}\,\rm km$.  Those parameters are closely matched by
our $1.33\,\Msun$ O/Ne WD model, which has
$T_{\rm eff} = 46 \,\rm kK$ and a radius of
$2,170\,\rm km$ at a cooling age of 70 Myr.
(For a C/O core composition, our models approximately match the best fit
parameters at mass of $1.345\,\Msun$ and a cooling age of 110 Myr.)
As can be seen in the isochrones shown in
Figure~\ref{fig:HR}, the best fit parameters place ZTF J190132.9+145808.7
in a region where neutrinos produced through the Urca process have accelerated
the cooling of the WD.

\begin{acknowledgements}

  We thank Evan Bauer for helpful comments and suggestions regarding
  WD cooling in \MESA, Ilaria Caiazzo, Jim Fuller, and Jeremy Heyl for
  comments on the manuscript and discussions about ZTF J190132.9+145808.7, and A.Y.~Potehkin for
  alerting us to the existence of updated plasmon neutrino
  emissivities.
  We thank the referee for a rapid and helpful report.
  J.S. is supported by the National Science Foundation through grant
  ACI-1663688 and via support for program number HST-GO-15864.005-A
  provided through a grant from the STScI under NASA contract
  NAS5-26555.
  We acknowledge use of the lux supercomputer at UC Santa Cruz, funded by NSF MRI grant AST 1828315.

\end{acknowledgements}

\clearpage

\bibliography{Urca_cooled_WDs.bib}

\end{document}